\documentstyle[12pt,epsf]{article}
\let\section=\subsection     \let\subsection=\subsubsection                

\def\htm{\hat{t}}

\def\la{\langle}
\def\ra{\rangle}

\def\k2av{\la k_T^2\ra}
\def\knav{\la k_T^2\ra}
\newcommand{\f}[2]{\frac{#1}{#2}} 

\begin{document}
\begin{center}
   {\large \bf EFFECTS OF DENSE MATTER ON HADRON}\\[2mm]
   {\large \bf PRODUCTION IN HEAVY-ION COLLISIONS
\renewcommand{\thefootnote}{\fnsymbol{footnote}}\setcounter{footnote}{2}%
\footnote{Contribution to Hirschegg 2000, Hadrons in Dense Matter, Jan. 17-21, 2000, Austria}}\\[5mm]
   G.~PAPP$^{1,2}$, G.~BARNAF\"OLDI$^3$, G.~FAI$^1$, P. L\'EVAI$^3$, Y.~ZHANG$^1$ \\[5mm]
   {\small \it  $^1$ CNR, Kent State University, Kent, OH, USA \\
                $^2$ HAS Research Group, ELTE, Budapest, Hungary \\
                $^3$ KFKI RMKI, P.O. Box 49, Budapest, Hungary \\[8mm] }
\end{center}

\begin{abstract}\noindent
The intrinsic transverse momentum distribution of partons in the
nucleon can be used to explain a large amount of high-$p_T$ hadron and
photon production data in high-energy nucleon-nucleon collisions at
energies $\sqrt{s} \approx 20$ to $1800$ GeV. However, proton-nucleus
experiments at energies $\approx 30$ GeV show an extra enhancement 
(Cronin effect) in the yield of photons and mesons
compared to a simple extrapolation of the proton-proton data. This
enhancement is due to the effect of the dense hadronic matter
encountered by the projectile proton in
the nuclear environment. We discuss the origin and the properties of 
the nuclear enhancement. 
\end{abstract}

\section{Introduction}

In hard particle production in proton-nucleus ($pA$) collisions,
the effect of the nuclear medium manifests itself in a cross section
enhancement at high transverse momentum relative to a simple 
superposition of the proton-proton ($pp$) production cross
section\cite{antreasyan79}. The purpose of this contribution is 
to study the Cronin effect, and to look for an
explanation in terms of the dense nuclear medium
that the projectile proton has to traverse in $pA$ collisions. 
The presence of the medium leads to an increase of the width
of the transverse momentum distribution of partons. The transverse 
momentum distribution of partons is recently also utilized 
in the description of dilepton data at SPS\cite{kampfer00}.  

\section{Transverse momentum distribution in $pp$ collisions}

The E609 collaboration extracted a width of 
$\langle k_T\rangle=(0.9 \pm 0.2)$ GeV/c
for the transverse momentum distribution of partons in the proton from
dijet events at 400 GeV/c\cite{corcoran91}. We 
use $\la k_T\ra$ as an energy dependent parameter
to obtain a satisfactory description of 
hard-particle production cross sections in the center-of-mass energy
range 18 GeV $\leq \sqrt{s} \leq$ 65 GeV in $pp$ collisions.
The transverse momentum
distribution enters in the perturbative QCD (pQCD) calculation of the 
hard-particle production cross section via the factorization-theorem 
based convolution
\begin{eqnarray}
\label{pQCD}
  E\frac{d\sigma^{pp}}{d^3p} &=&
        \sum_{abcd}\!  \int\!dx_1 dx_2\ d^2\!k_{T1}\ d^2\!k_{T2}\
        f_{a/p}(x_1,Q^2) g_1({\vec k}_{T1})\
        f_{b/p}(x_2,Q^2) g_2({\vec k}_{T2})\ \nonumber \\
          && \ \ \ \    \frac{d\sigma}{d\htm}(ab\to c d)\,
   \frac{D_{c}(z_c,{\widehat Q}^2)}{\pi z_c} \ \ \  , 
\end{eqnarray}
where the connection to the observed incoming and outgoing 
particles is established through the
parton distribution functions (PDFs) $f(x,Q^2)$ and 
fragmentation functions (FFs) $D(z,{\widehat Q}^2)$\cite{Field89}.
The value of $z_c$ is fixed by energy-momentum conservation.
The transverse momentum distribution is denoted by
$g({\vec k}_{T})$, and as an approximation,
is taken to be a Gaussian:
$g({\vec k}_T) = \exp(-k_T^2/\langle k_T^2 \rangle)/
{\pi \langle k_T^2 \rangle}$.
Here, $\langle k_T^2 \rangle$ is the 2-dimensional width of the $k_T$
distribution and it is related to the average transverse momentum of one parton
as $\langle k_T^2 \rangle = 4 \langle k_T \rangle^2 /\pi$. 

The cross section of the hard subprocesses in Eq. (\ref{pQCD}) is to be 
calculated in pQCD. In principle, the order in the strong 
coupling constant, to which this is done, is a matter of decision based 
on the convergence of the series. At next-to-leading order
(NLO), the cross section can be viewed as 
\begin{equation}
\label{Kdef}
\f{d\sigma^{NLO}}{d\htm} = \f{d\sigma^{NLO}_{Born}}{d\htm} +
   \f{d\sigma^{NLO}_{corr}}{d\htm} = K  \f{d\sigma^{NLO}_{Born}}{d\htm}
\;\;  \, \\,
\end{equation}    
where the factor $K$ may depend on both, the center-of-mass energy
and the transverse momentum, and represents an approximate way
to take into account corrections beyond the Born term. 
For sufficiently high energies and transverse momenta $K=const.$ is a good
approximation\cite{wong98}. We adapted $K=2$ in the present 
calculation. For the PDFs we used the MRST98 set\cite{MRST98}, which
is a NLO parameterization and incorporates an intrinsic $k_T$. The scales are 
fixed, $\Lambda_{\overline {MS}}(n_f=4) = 300$ MeV and $Q=\widehat{Q}=p_T/2$. 
An NLO parameterization was used for the FFs, too\cite{BKK}, thus 
assuring the consistency of Eq. (\ref{pQCD}). The energy dependence of
$\k2av$ and the quality of the agreement with hard $\pi^0$ and $\gamma$
production data in $pp$ collisions is displayed in Ref. \cite{plf00}.
Note that a satisfactory description of the pion spectra does not 
appear possible in NLO without transverse-momentum smearing\cite{Narancs}.  

\section{Enhanced transverse momentum in $pA$ collisions}

The Cronin effect in $pA$ collisions can be understood
phenomenologically\cite{wang9798}
in terms of an increased $\k2av$ according to 
\begin{equation}
\label{ktbroadpA}
\knav_{pA} = \knav_{pp} + C \cdot h_{pA}(b) \ . 
\end{equation}
Here, $h_{pA}(b)$ describes the number of 
nucleon-nucleon collisions at impact parameter $b$,
which contribute to the transverse momentum broadening of the
partons in the projectile. The particle-producing hard 
collision is clearly not counted, but it is also not obvious that
all other possible collisions contribute. We therefore use the term
{\it effective} to distinguish the transverse-momentum 
broadening collisions. Each  effective collision imparts
a transverse momentum squared $C$ on average. 

If it is assumed that the projectile proton suffers soft collisions, 
which do not change the projectiles identity as a proton, but
enhance the width of the transverse momentum distribution of its
partons via random interactions, each imparting some transverse 
momentum, then all collisions of the projectile 
at impact parameter $b$ are effective. This can be expressed by
using $h^{all}_{pA}(b)=\nu_A(b) - 1$,
where $\nu_A(b) = \sigma_{NN} \ t_A(b)$ 
is the collision number at impact parameter $b$, with $\sigma_{NN}$
being the total nucleon-nucleon cross section. 
However, using this prescription,
the coefficient $C$ turns out to be target dependent\cite{plf00}. 

Furthermore, it is necessary to use values as high as $C \approx 1$ GeV$^2$
with this prescription, no longer characteristic of soft 
physics. Thus it is natural to assume that the projectile proton 
suffers a more drastic change while traveling through the dense 
matter of the target nucleus. In particular, we assume that the 
projectile proton will break up in
a hard or semi-hard collision. As a consequence of this violent 
event, partonic degrees of freedom become appropriate.
The enhanced width of the transverse momentum distribution reflects this
change of the physically relevant degrees of freedom, and gets
``measured'' in the collision that breaks the proton apart.  

We studied scenarios where
only a fraction of the possible $\nu_A(b)-1$ collisions is effective
in the sense of nucleon-nucleon collisions. In the limiting case,
referred to as the {\it saturating} prescription, we 
allow at most one effective collision using a 
smoothed step function $h^{sat}_{pA}(b)$, defined as

\begin{equation}
\label{hsat}
 h^{sat}_{pA}(b)  = \left\{ \begin{array}{ll}
           0  & \mbox{ if \ $\nu_A(b) < 1$ } \\
  \nu_A(b)-1  & \mbox{ if \ $1 \leq \nu_A(b) < 2$ } \ \ \ . \\
           1  & \mbox{ if \ $2 \leq \nu_A(b)$  } 
           \end{array} \right. 
\end{equation}
The saturated Cronin factor is denoted by  $ C^{sat}$. 
By the saturating prescription we limit 
our consideration to at most one {\it effective}
transverse momentum broadening collision in the above sense.

\begin{figure}[tbp]
\centerline{\epsfxsize=0.95\textwidth \epsfbox{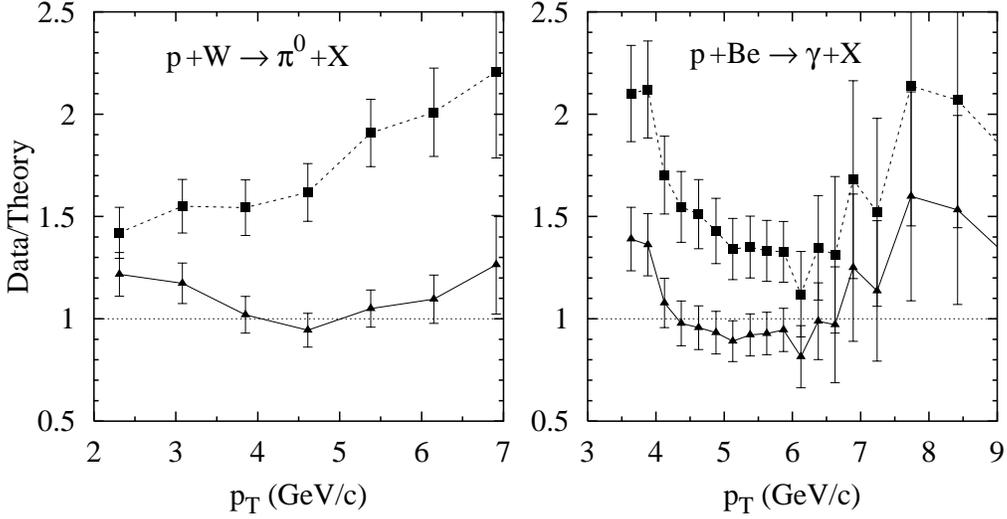}}
\caption{\small {\bf Left panel}: Data to theory ratio on a linear scale for the 
$pW \rightarrow \pi^0 X$ reaction at $\sqrt{s}=27.4$ GeV, data from
\cite{antreasyan79}. We show $C^{sat}=1.2$ GeV$^2$ (full line) and
$C^{sat}=0$ (dashed line).
{\bf Right panel}: Data to theory ratio for the 
$pBe \rightarrow \gamma X$ reaction at $\sqrt{s}=31.6$ GeV, data from 
\cite{apanasevich98}.
Solid line indicates $C^{sat}=1.2$ GeV$^2$, dashed line means $C^{sat}=0$.
}
\label{fig-rat}
\end{figure}
 
The left panel of Fig 1. shows the data/theory ratio on a linear scale for the
$pW \rightarrow \pi^0 X$ reaction with (full line) and without (dashed line) the 
saturated Cronin enhancement, using the value $C^{sat}=1.2$ GeV$^2$, with the
data from Ref. \cite{antreasyan79}.  
In the left panel we illustrate that the same value of $C_{sat}$
provides a satisfactory description of recent photon production data 
in p + Be collisions in the same energy region\cite{apanasevich98}.
Since the effect of the nuclear medium on the projectile proton should not 
depend on the outgoing particle, we consider the common value of $C_{sat}$
a strength of our model. 

Our $C^{sat}$ is determined at $\sqrt{s} \approx 30$ GeV. Its value may
of course depend on the center-of-mass energy. This dependence needs to 
be mapped out in a wide energy range up to $\sqrt{s}=200$ GeV in order 
to make the model useful for RHIC predictions. On the other hand, 
we expect $C^{sat}$ to be independent of the target and of the produced hard 
particle. Systematic experimental studies to test these predictions
would be welcome.

\section{Summary}

We reported on a phenomenological model to understand the Cronin effect.
The enhancement of hard particle production cross sections in $pA$ 
collisions relative to scaled $pp$ results was connected to an
increase in the width of the transverse momentum distribution of 
the projectile as it traverses the dense nuclear medium of the target.
Physically rather different pictures for the enhancement of the transverse 
momentum width were studied. The extreme saturating prescription gives 
reasonable agreement with available data at the present level of the
calculation. 

\section{Acknowledgement}

This work was supported in part by DOE grant DE-FG02-86ER40251 and 
Hungarian grant OTKA-T032796. Partial
support from the Domus Hungarica program of the Hungarian Academy of Sciences 
is gratefully acknowledged.

\end{document}